\documentclass[sigplan,screen]{acmart}
\AtBeginDocument{%
  }

\copyrightyear{2025}
\acmYear{2025}
\setcopyright{acmlicensed}\acmConference[ASPLOS '25]{Proceedings of the 30th ACM International Conference on Architectural Support for Programming Languages and Operating Systems, Volume 3}{March 30-April 3, 2025}{Rotterdam, Netherlands}
\acmBooktitle{Proceedings of the 30th ACM International Conference on Architectural Support for Programming Languages and Operating Systems, Volume 3 (ASPLOS '25),
March 30-April 3, 2025, Rotterdam, Netherlands}
\acmDOI{10.1145/3676642.3736120}
\acmISBN{979-8-4007-1080-3/2025/03}

\begin{document}


\newcommand{\archname}{PUSHtap}
\title{\archname{}: PIM-based In-Memory HTAP with Unified Data Storage Format}

\author{Yilong Zhao}
\affiliation{%
  \institution{Shanghai Jiao Tong University}
  \city{Shanghai}
  \country{China}}
\affiliation{%
  \institution{Shanghai Qi Zhi Institute}
  \city{Shanghai}
  \country{China}}
\email{sjtuzyl@sjtu.edu.cn}
\orcid{0000-0001-8291-6896}

\author{Mingyu Gao}
\affiliation{%
  \institution{Tsinghua University}
  \city{Beijing}
  \country{China}}
\affiliation{%
  \institution{Shanghai Qi Zhi Institute}
  \city{Shanghai}
  \country{China}}
\email{gaomy@tsinghua.edu.cn}
\orcid{0000-0001-8433-7281}

\author{Huanchen Zhang}
\affiliation{%
  \institution{Tsinghua University}
  \city{Beijing}
  \country{China}
}
\affiliation{%
  \institution{Shanghai Qi Zhi Institute}
  \city{Shanghai}
  \country{China}
}
\email{huanchen@tsinghua.edu.cn}
\orcid{0009-0001-4821-1558}

\author{Fangxin Liu}
\authornote{This work was partially supported by the National Key Research and Development Program of China (2024YFE0204300), National Natural Science Foundation of China (Grant No.62402311), and Natural Science Foundation of Shanghai (Grant No.24ZR1433700). Fangxin Liu and Li Jiang are the corresponding authors.}
\affiliation{%
 \institution{Shanghai Jiao Tong University}
 \city{Shanghai}
 \country{China}
}
\affiliation{%
  \institution{Shanghai Qi Zhi Institute}
  \city{Shanghai}
  \country{China}
}
\orcid{0000-0002-8769-293X}
\email{liufangxin@sjtu.edu.cn}

\author{Gongye Chen}
\affiliation{%
  \institution{Shanghai Jiao Tong University}
  \city{Shanghai}
  \country{China}
}
\email{gongye_chen@sjtu.edu.cn}
\orcid{0009-0005-9944-414X}

\author{He Xian}
\affiliation{%
  \institution{Shanghai Qi Zhi Institute}
  \city{Shanghai}
  \country{China}
}
\email{51265900021@stu.ecnu.edu.cn}
\orcid{0009-0008-2099-4109}

\author{Haibing Guan}
\affiliation{%
  \institution{Shanghai Jiao Tong University}
  \city{Shanghai}
  \country{China}
}
\orcid{0000-0002-4714-7400}
\email{hbguan@sjtu.edu.cn}

\author{Li Jiang}
\authornotemark[1]
\affiliation{%
  \institution{Shanghai Jiao Tong University}
  \city{Shanghai}
  \country{China}
}
\affiliation{%
  \institution{Shanghai Qi Zhi Institute}
  \city{Shanghai}
  \country{China}
}
\orcid{0000-0002-7353-8798}
\email{jiangli@cs.sjtu.edu.cn}

\renewcommand{\shortauthors}{Yilong Zhao et al.} 

\begin{abstract}
Hybrid transaction/analytical processing (HTAP) is an emerging database paradigm that supports both online transaction processing (OLTP) and online analytical processing (OLAP) workloads.
Computing-intensive OLTP operations, involving row-wise data manipulation, are suitable for row-store format.
In contrast, memory-intensive OLAP operations, which are column-centric, benefit from column-store format.
This \emph{data-format dilemma} prevents HTAP systems from concurrently achieving three design goals: performance isolation, data freshness, and workload-specific optimization.
Another background technology is Processing-in-Memory (PIM), which integrates computing units (PIM units) inside DRAM memory devices to accelerate memory-intensive workloads, including OLAP.

Our key insight is to combine the interleaved CPU access and localized PIM unit access to provide two-dimensional access to address the data format contradictions inherent in HTAP.
First, we propose a unified data storage format with novel data alignment and placement techniques to optimize the effective bandwidth of CPUs and PIM units and exploit the PIM's parallelism.
Second, we implement the multi-version concurrency control (MVCC) essential for single-instance HTAP.
Third, we extend the commercial PIM architecture to support the OLAP operations and concurrent access from PIM and CPU.
Experiments show that \archname{} can achieve 3.4\texttimes{}/4.4\texttimes{} OLAP/OLTP throughput improvement compared to multi-instance PIM-based design.
\end{abstract}

\ccsdesc[500]{Computer systems organization~Heterogeneous (hybrid) systems}
\ccsdesc[500]{Information systems~Database design and models}

\keywords{Processing-in-Memory (PIM), Hybrid Transactional/Analytical Processing (HTAP), DRAM, Unified Data Format}


\maketitle

\section{Introduction}
\label{sec:introduction}

\begin{figure}
    \centering
    \includegraphics[width=\linewidth]{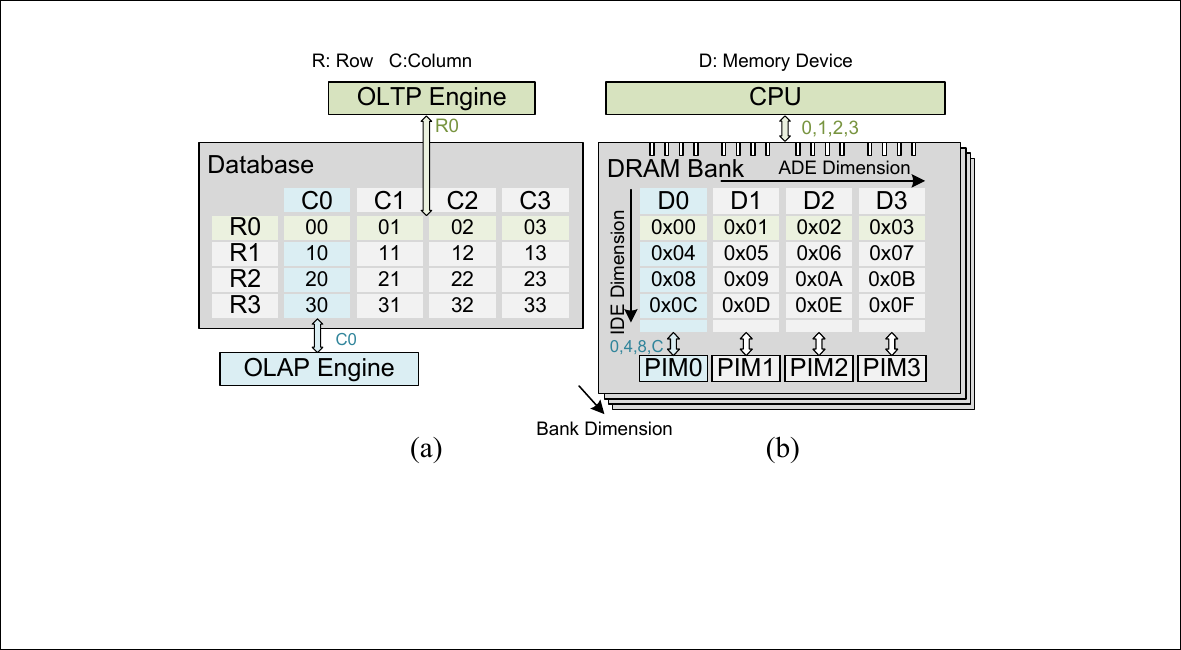}
    \vspace{-15pt}
    \caption{(a) HTAP system: OLTP and OLAP engines process rows and columns, respectively. (b) DRAM-based PIM architecture: CPU interleaves data across memory devices (ADE), and PIM units access data inside the device (IDE).}
    \label{fig:interleaving&htap}
    \vspace{-5pt}
    \Description{This figure shows the similarities between HTAP and PIM architecture. In subfigure (a), we present the HTAP system, where the OLTP engine processes rows and the OLAP engine processes columns of a table. In subfigure (b), the DRAM-based PIM architecture interleaves data across memory devices (ADE), and PIM units access data inside the device (IDE). The ADE dimension is similar to the row dimension in HTAP, while the IDE dimension is similar to the column dimension in HTAP.}
\end{figure}

Hybrid transaction/analytical processing (HTAP) is an emerging processing architecture that allows one database system to support two processing sets: online transaction processing (OLTP) and online analytical processing (OLAP) \cite{first-htap2014hybrid}.
There has been extensive research and design for HTAP \cite{Hyper5767867, C-store10.1145/3226595.3226638,singlestore, Polynesia9835628, OracleDF7113373, SAPHANA10.1145/2213836.2213946}.
As shown in \figurename{}~\ref{fig:interleaving&htap} (a), the OLTP engine processes transactions, which are single-record operations on rows, including read, insert, update, and delete.
On the contrary, the OLAP engine processes analytical queries to solve multidimensional analysis problems on columns.
OLTP includes read and write operations, whereas all OLAP operations are read operations.
An ideal HTAP system should meet the following three design goals \cite{BatchDB10.1145/3035918.3035959, Polynesia9835628, HTAPReview10.1145/3514221.3522565}.
(1) \emph{Workload-specific optimizations}: optimized performance for both OLTP and OLAP workloads.
(2) \emph{Performance isolation}: limited performance degradation for concurrent execution of transactions and analytical queries.
(3) \emph{Data refreshes}: analytical queries need up-to-date data. 
However, current HTAP systems cannot fulfill three goals simultaneously due to the diverse data storage formats required by OLTP and OLAP, as analyzed below. 

\begin{figure}
    \centering
    \includegraphics[width=0.95\linewidth]{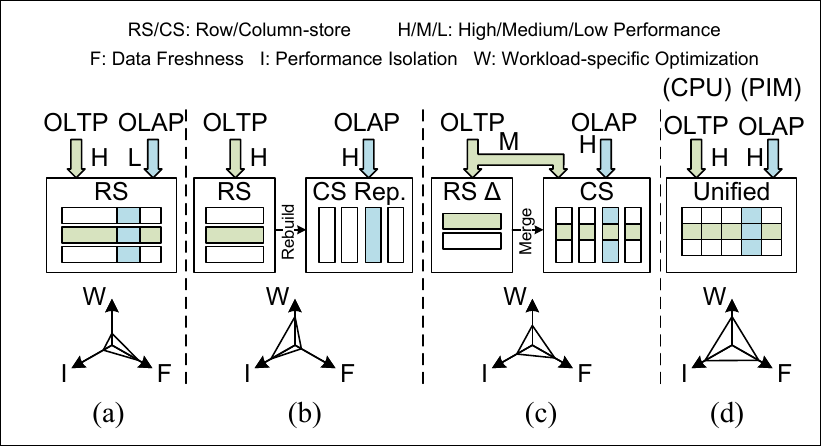}
    \vspace{-10pt}
    \caption{Different HTAP data format design. (a) Single-instance design with single data format. (b) Multiple-instance design with different data formats. (c) Single-instance design with mixed data format. (d) \archname{} with single-instance and unified data format.}
    \vspace{-5pt}
    \label{fig:previous_work}
    \Description{In this figure, we summarize three predominantly adopted data formats in current HTAP systems. In subfigure (a), we present the single-instance design with a single data format. Both OLTP and OLAP engines process the same data format (row-store or column-store). This format can achieve high data freshness by compromising the workload-specific performance and performance isolation. In subfigure (b), the multi-instance design with different data formats is presented. Both OLTP and OLAP engines process on their preferred data formats, and column-store instances need to be rebuilt from the latest row-store data after a large interval. This design can achieve optimal workload-specific performance and high-performance isolation, but has low data freshness. In subfigure (c), the single-instance design with mixed data format is presented. The database is primarily column-store, and the row-store delta is used to store the newest version of rows updated by transactions. This design can achieve high data freshness, but has medium workload-specific performance and medium performance isolation. In the above three designs, the three design goals cannot be fulfilled simultaneously, becoming an impossible triangle. In subfigure (d), the \archname{} with single-instance and unified data format is adopted. Both OLAP and OLTP engines process the same data format, which can achieve all three design goals.}
\end{figure}

\figurename{}~\ref{fig:previous_work} (a,b,c) summarizes three predominantly adopted data formats in current HTAP systems.
(1) \emph{Single-instance with single data format}, either column or row store (\figurename{}~\ref{fig:previous_work}~(a)) \cite{Hyper5767867, C-store10.1145/3226595.3226638}.
This design can achieve high data freshness because the OLAP engine is always visible to the latest data that is updated by the OLTP engine.
However, a single data format results in sub-optimal workload-specific performance.
For example, performing analytical queries on row-store has 50\% performance degradation when over 95\% of the accessed data are unused; while transactions on column-store have more than 20\% degradation \cite{cs_testDBLP:journals/corr/abs-2109-14349}.
(2) \emph{Multi-instance design}~\cite{singlestore, Polynesia9835628}.
As shown in~\figurename{}~\ref{fig:previous_work}~(b), both transactions and analytical queries process on their preferred data formats, and thereby can obtain optimal workload-specific performance and high-performance isolation.
However, the HTAP system must rebuild the column-store replication from the latest row-store data to ensure data freshness, which requires a significant latency.
The rebuilding cannot be executed frequently, leading to low data freshness \cite{HTAPReview10.1145/3514221.3522565}.
(3) \emph{Single-instance design with mixed data format} \cite{OracleDF7113373, SAPHANA10.1145/2213836.2213946}.
This design is composed of a primary column-store data and a row-store delta, as shown in~\figurename{}~\ref{fig:previous_work}~(c).
The row-store delta maintains the newest version of rows updated by transactions.
OLTP and OLAP engines must scan the two format regions to acquire the complete data.
For example, the OLAP engine first scans the column-store data with high bandwidth, then acquires the newest data version by scanning the row-store delta with low bandwidth, and finally merges the data.
This approach ensures the freshness of the data but compromises the performance isolation of OLAP and OLTP.

Processing-in-memory (PIM) integrates computing units, i.e., PIM units, inside memory devices to accelerate memory-intensive workloads \cite{UPMEM8875680, AiMHW9731711, AiMSW9895629, HBM-PIM-SW9499894, HBM-PIM-HW9365862}.
PIM units directly access data on their own devices, bypassing the prolonged and limited memory bus to utilize the internal memory bandwidth better.
For example, a commercial PIM architecture can achieve over 3.3\texttimes{} bandwidth improvement and 10\texttimes{} access energy reduction \cite{UPMEM8875680}. 
One key limitation of PIM is its localized access, preventing efficient access to remote data residing on other devices.
For example, in \figurename~\ref{fig:interleaving&htap}~(b), PIM 0 can directly access data \texttt{0x00,04,08,0C} in the device \texttt{D0} within several nanoseconds. 
While for those in device~\texttt{D1,D2,D3}, they need CPU to move the data to their local device, costing around 0.2$\mu$s due to the mode-switch overhead \cite{UPMEM8875680}.

Memory interleaving is ubiquitously employed to improve CPU's memory bandwidth. 
\figurename~\ref{fig:interleaving&htap}~(b) presents a memory space interleaved to four devices.
Contiguous data blocks, \texttt{0x00-04}, are split and mapped to the same location of devices, respectively.
CPU can access four blocks in parallel within a single memory access, thereby maximizing the memory bandwidth.

This work advocates combining PIM's localized access and CPU's interleaved access to provide a new perspective for two-dimensional access to every memory bank: the CPU accesses data \emph{across devices} (ADE) in parallel; massive PIM units simultaneously access local data \emph{inside devices} (IDE) with low latency.
We can map the HTAP row to the ADE dimension and the column to the IDE dimension.
Accordingly, the CPU works as an OLTP engine, and PIM units work as an OLAP engine. 
Other memory hierarchies, e.g., bank~(can scale to rank and channel), form the third access dimension for database scalability.

In this work, we propose \archname{}, a PIM-based single-instance in-memory HTAP with the unified data storage format as outlined earlier, as shown in \figureautorefname{}~\ref{fig:previous_work} (d).
OLTP and OLAP engines can achieve optimal performance and data freshness through two-dimensional access on the instance.
The contribution of this paper is as follows:
\begin{itemize}
    \item \textbf{Unified HTAP Architecture:}
    We propose \archname{}, a PIM-based single-instance HTAP architecture with a unified data storage format. 
    The key insight behind \archname{} is the combination of interleaved CPU access and localized PIM unit access to address the data format contradictions inherent in HTAP.
    We introduce a data layout algorithm specifically designed for \archname{} to enhance effective bandwidth across databases with varying column widths.
    \item \textbf{Concurrency Control Operations:} 
    We design the multi-version concurrency control (MVCC) and the corresponding snapshot operations for \archname{}.
    These features are crucial for a single-instance database environment to minimize data transfer between CPU and PIM units, thus optimizing overall system efficiency.
    \item \textbf{Architecture Support:}
    We extend the memory controller of DRAM-based general-purpose PIM (UPMEM-like) \cite{UPMEM8875680}, with a hardware interface to support the OLAP operations and the concurrent access by CPU and PIM units, which is crucial to \archname{}.
\end{itemize}
\section{Background}
\label{sec:background}

\subsection{DRAM-based PIM }
\label{sec:background:DRAM-PIM}

DRAM-based PIM architectures are proposed to accelerate memory-intensive applications \cite{HBM-PIM-HW9365862,UPMEM8875680,AiMHW9731711}.
\cite{UPMEM8875680} is a representative commercial general-purpose PIM architecture.
A PIM unit, along with two scratchpad SRAMs, WRAM (buffer operands), and IRAM (buffer instructions), is integrated into each memory bank.
A PIM interface is added to each rank to facilitate CPU control over PIM units.
Memory-intensive operations are offloaded to PIM units to better utilize the internal bandwidth of DRAM.

A general-purpose PIM is designed with two modes: CPU mode and PIM mode \cite{UPMEM8875680}.
In CPU mode, CPU accesses DRAM banks as conventional memory.
In PIM mode, PIM units fully control data flow and computation, and DRAM banks are locked to prevent CPU access.
When switching from CPU mode to PIM mode, CPU needs to send messages to all the PIM units to hand over the bank access control and invoke the PIM units.
After that, CPU polls the PIM units until all the PIM units are finished.
As there are thousands of PIM units in a server (4 channels, 8 DDR4 DIMMs \cite{UPMEM8875680}), it takes \textbf{tens of microseconds} to invoke and poll them.
With these offloading overheads, PIM is beneficial only for coarse-grained PIM tasks \cite{UPMEM-Benchmarking9771457}.
The two-mode design is a general design for PIM, and it is also adopted by other PIM architectures, such as HBM-PIM \cite{HBM-PIM-HW9365862}.

\subsection{PIM-based HTAP Systems}
\label{sec:background:htap_pim}

OLAP operations are performed on massive columns and are memory-intensive \cite{mondrian8192508, Polynesia9835628,memoryintensiveanalys10.1145/2254756.2254766,kim2023darwin}.
OLAP operations are proposed to be accelerated by PIM. 
This approach prevents the lengthy data movement between memory and CPU.
Moderian \cite{mondrian8192508} proposes an execution model for analytical queries on HMC-based PIM.
The data are divided and moved to the destination vault for the following computation during each query, introducing additional memory movement and violating the PIM's design principle.
Polynesia \cite{Polynesia9835628} adopts a multi-instance design on HBM, with instances stored in both CPU and PIM memory space.
Their approach is still the same as the multi-instance design with mixed data formats (\figureautorefname{}~\ref{fig:previous_work} (b)) and requires rebuilding the column-store instance through logs.
The rebuilding involves transferring both transaction logs and new-versioned data to the PIM memory space.
Despite the integration of supplementary hardware for merging the log, this procedure still negatively impacts analytical query performance.
Specifically, with 8M transactions, the query time is increased by 18\% according to their experimental results (\figureautorefname{}~9 (b) of \cite{Polynesia9835628}).
In summary, these approaches remain limited to existing data format designs and do not fully exploit PIM's advantage of minimizing data movement.

\subsection{Database Concurrency Control}
\label{sec:background:htap_storage}

MVCC is a widely used concurrency control method for single-instance database systems.
A \emph{metadata} is maintained for every row to facilitate transactions and analytical queries \cite{MVCC10.14778/3551793.3551832, mvccforlongtransaction10.1145/3318464.3389714,htap-mvcc10.1145/3514221.3526135}.
The metadata contains three fields: a \emph{read timestamp}, a \emph{write timestamp}, and a \emph{pointer}.
The write timestamp records the transaction that creates the version, and the read timestamp records the transaction of the most recent read.
The pointer of this version points to the previous version of this row.
Upcoming transactions to the same row may form a \emph{version chain}.
Before analytical queries, the timestamps and the pointers are scanned to create a \emph{snapshot} to summarize the visible version of data for OLAP engine processing \cite{Polynesia9835628, OracleDF7113373, SAPHANA10.1145/2213836.2213946}.
The snapshot is a collection of data pointers that have a consistent version.
The analytical query is processed on the snapshot to maintain the version's consistency and make sure they do not scan the new versions created by concurrently issued transactions.
For large-scale databases, the snapshot is continuously updated based on newly arriving transactions rather than being rebuilt from scratch each time \cite{dbx100010.14778/2735508.2735511}.
Moreover, the memory space becomes fragmented because MVCC frequently allocates new memory for new-versioned rows and releases memory for stale rows.
This usually reduces the memory access efficiency \cite{sql-defragmentation} because memory \emph{defragmentation} is processed periodically in the database with MVCC.

\section{Challenges}
\label{sec:motivation}

In this work, we propose \archname{}, motivated by the potential to optimize HTAP through integrated interleaved CPU and localized PIM unit access.
This approach brings new opportunities to achieve all three design goals using a unified data format and single-instance design.
However, we still face some challenges as presented below.

\textbf{(1) Data Format Challenge---Data Alignment for Bandwidth Effectiveness.}
We refer to the data size of each element on ADE dimension and IDE dimension's intersection as \emph{interleave granularity}, shown as the data blocks in \figurename{}~\ref{fig:interleaving&htap} (b), indicating the minimum data size the CPU and PIM can read from a memory sub-module during each access. 
The interleave granularity is fixed to 8B in DIMM-based PIM due to the specification of the protocol. 

\begin{figure*}
    \centering
    \includegraphics[width=\linewidth]{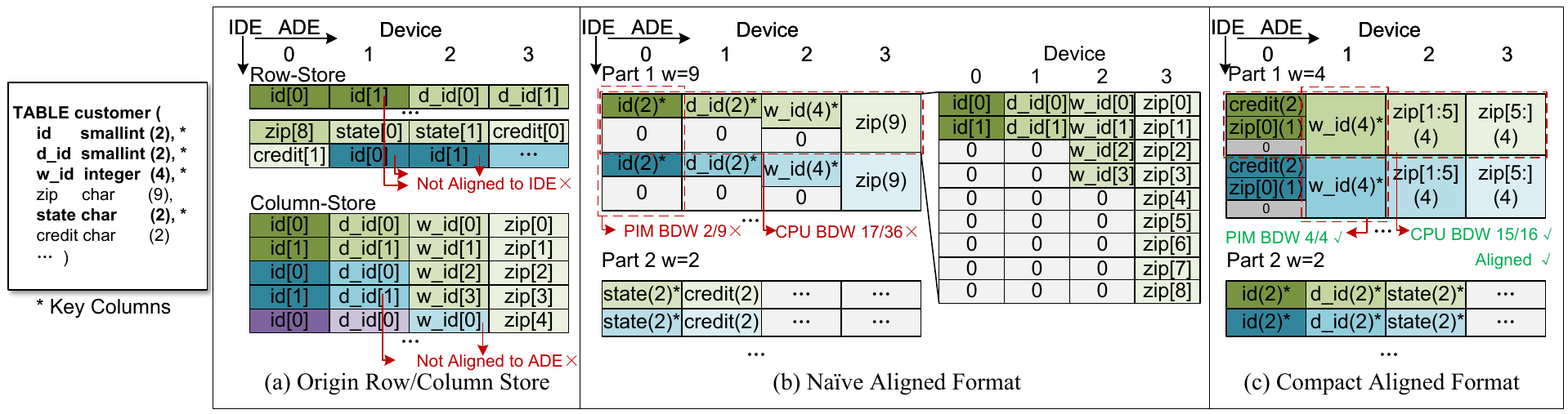}
    \vspace{-10pt}
    \caption{(a) Row-store format in current CPU-based HTAP and column-store format in current PIM-based HTAP. Different colors represent different rows. (b) A na\"ive aligned format for a database with various column widths. (c) The proposed compact aligned format achieves both alignment and high effective bandwidth. ($w$: Column width)}
    \label{fig:data-format}
    \Description{In this figure, we present the data format of current HTAP systems, naive and compact aligned format. This figure's example is based on a table in CH-Benchmark whose column width varies from 2 to 9 bytes. In subfigure (a), we present that both row-store and column-store formats cannot satisfy the requirement to align the data to the ADE and IDE dimensions simultaneously. In subfigure (b), a na\"ive aligned format for a database with various column widths is presented, by padding excessive dummy `0's to align all the columns to the widest column, i.e., 9 bytes. In subfigure (c), the proposed compact aligned format achieves alignment and inserts very few dummy `0's, which can achieve high effective bandwidth.}
\end{figure*}

According to our insight, the rows and columns are aligned to the ADE and IDE dimensions, respectively.
However, with variable-sized column width, the traditional row-store or column-store cannot satisfy the requirement.
For example, in table CUSTOMER of CH-Benchmark \cite{CH-benchmark10.1145/1988842.1988850, tpcc, tpch}, the column width varies from 2 to 9 bytes.
The conventional row-/column-stored format is shown in \figurename{}~\ref{fig:data-format} (a).
If the table is row-stored, the columns are not aligned to IDE because the row size is not a multiple of the device number.
For example, Column \texttt{id} of the first row is mapped to devices 0 and 1, while that of the second row is mapped to devices 1 and 2.
On the other hand, if the table is column-stored as in \cite{C-store10.1145/3226595.3226638}, rows are not aligned to the ADE dimension.
Elements of a row, e.g., \texttt{id} and \texttt {zip} of the second row, are distributed across different cache lines.
This data format costs the CPU multiple bursts to access a row during transactions, resulting in low effective bandwidth.

In \sectionautorefname{}~\ref{sec:layout:alignment}, we propose a novel compact aligned format to maintain hardware alignment while optimizing bandwidth usage and preserving data integrity.

\textbf{(2) Data Format Challenge---PIM Parallelism.} 
Analytical queries are executed on only several columns.
As presented in  \figureautorefname{}~\ref{fig:block_rotation} (a) of \sectionautorefname{}~\ref{sec:layout:parallelism},
mapping a whole column to the IDE dimension actually worsens the parallelism of PIM units.
We thus propose a block-circulant format to fully utilize the parallelism of PIM units to optimize analytical queries' performance in \sectionautorefname{}~\ref{sec:layout:parallelism}.

\textbf{(3) Data Movement Challenge in MVCC.}
In single-instance database design with MVCC, snapshot and defragmentation are two necessary operations.
Unlike CPU-based HTAP systems, the snapshot should be transferred to PIM units in \archname{}.
This requires encoding the snapshot to minimize the data transfer.
Moreover, \archname{} focuses on compact data placement to ensure effective bandwidth.
The fragmentation caused by MVCC brings severe performance degradation.
We need to reduce the defragmentation overhead by transferring less data and utilizing the large PIM bandwidth to allow more frequent defragmentation execution.
The snapshot and defragmentation operation designed for \archname{} is presented in \sectionautorefname{}~\ref{sec:operations}.

\textbf{(4) Hardware Challenge.}
Current PIM designs presented in \sectionautorefname{}~\ref{sec:background:DRAM-PIM} cannot satisfy \archname{}'s requirements in the following two aspects.
Firstly, in a single-instance HTAP database, the OLTP and OLAP engines process the same data instance concurrently, requiring CPU and PIM units to concurrently access the same banks.
However, current PIM architectures only benefit from coarse-grained PIM tasks due to the significant offloading overhead.
During the coarse-grained PIM task, PIM occupies the bank for over seconds, whether or not it is accessing the banks.
This results in long transaction latency and cannot fulfill many scenarios:
Many databases are timely databases and require a microsecond-level delay\cite{htapscale10.14778/2824032.2824069}.
In \sectionautorefname{}~\ref{sec:architecture}, we design the hardware and software interface for OLAP operations by automatically controlling the PIM units.

\section{The Unified Data Format}
\label{sec:layout}

\begin{figure}
    \centering
    \includegraphics[width=0.75\linewidth]{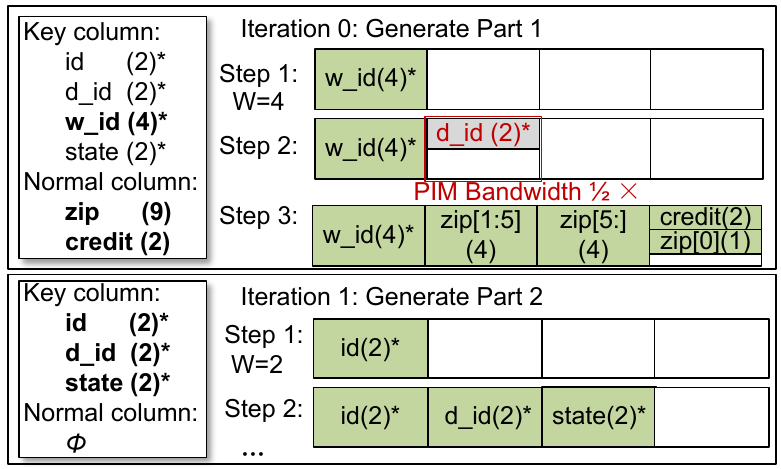}
    \caption{Generating the compact aligned format.}
    \label{fig:formatting-algirithm}
    \Description{This figure shows the compact aligned format generation algorithm. The algorithm is based on the bin-packing algorithm. The algorithm starts with an empty part and selects the widest key column \texttt{w\_id} and places it in the first device of this part. Then, the algorithm selects the rest key columns whose width is larger than $th$ times \mathrm{4B}, whose result is 3B, and places them in the second device of this part. Finally, the algorithm fills the rest of the bytes with the normal columns and places them in the following devices.}
    \vspace{-5pt}
\end{figure}

In this section, we present the unified data format of \archname{} that optimizes the performance of both row and column-wise operations in HTAP.
We first present a \emph{compact aligned} format method for a database with variable column widths to fit the fixed interleave granularity.

\subsection{Aligned Data Format}
\label{sec:layout:alignment}

\subsubsection{Na\"ive Aligned Format}
\label{sec:layout:alignment:naive}

One na\"ive aligned format that all the rows and columns are aligned with ADE and IDE, respectively, is presented in \figurename{}~\ref{fig:data-format} (b).
Since our example has only four devices, the table is divided into two parts.
The columns \texttt{id}, \texttt{d\_id}, \texttt{w\_id}, and \texttt{zip} are stored in the first part, while \texttt{state} and \texttt{credit} are stored in the second part.
The two parts are mapped to different memory channels so the CPU can access them in parallel.
The widest column is \texttt{zip} (9 bytes), and all other columns are aligned to this width by padding zeros for a na\"ive alignment.
In the following discussion, we refer to the width of the widest column as \emph{row width}.
Part 1's row width is 9, and Part 2's is 2. 

This na\"ive format wastes not only memory capacity but also CPU's and PIM's bandwidth.
When the CPU reads one row in part 1, it reads 9 bytes from each device.
However, only 17 of the $4\times 9$ bytes contain actual data.
The same situation also occurs when PIM units access DRAM devices.
When the PIM unit processes column \texttt{id}, 8-byte contiguous data is loaded to the SRAM buffer in each access because the data wire is 64-bit wide \cite{UPMEM8875680}.
Only 2 out of 8 bytes contain actual data, wasting 75\% of the PIM bandwidth.

\subsubsection{Compact Aligned Format}
\label{sec:layout:alignment:compact}

To further improve the effective bandwidth, we propose a compact aligned format for \archname{} to improve the bandwidth efficiency of PIM and CPU.
We can take advantage of the following two observations.
First, bytes in a row can be reordered, and columns of similar widths can be mapped to the same part, reducing zero padding.
Second, some columns are not scanned in any frequent analytical queries. These OLAP-free columns can be split and mapped to multiple devices, leading to a smaller row width and reducing the number of dummy `0's padded.

In the following discussion, we denote the columns scanned by analytical queries as \emph{key columns} and other columns as \emph{normal columns}.
For example, column \texttt{id} is scanned by Query 3.
While column \texttt{zip} is not operated by any query in CH-benchmark \cite{CH-benchmark10.1145/1988842.1988850,tpch}.
Therefore, \texttt{id} is a key column, while \texttt{zip} is a normal column.

Based on the above observations, we present a strategy based on the bin-packing algorithm to generate a compact aligned format, as shown in \figureautorefname{}~\ref{fig:formatting-algirithm}.
The format can keep the alignment to ADE and IDE dimensions and simultaneously achieve a high effective bandwidth.
We introduce a hyperparameter, the threshold $th$.
In our example, we set $th=3/4$.
In each iteration, we generate the format of one part of the table.
Firstly, we start with an empty part and select the widest key column \texttt{w\_id} and place it in the first device of this part.
Therefore, this part's row width is determined as 4 bytes.
Then in the second step, we select from the rest key columns whose width $\geq th\cdot \mathrm{4B}$ (3B).
In our example, there is no rest key column satisfying the condition.
If we place a key column with a too-small width, for example, \texttt{d\_id} of 2B width, 50\% of PIM bandwidth is wasted when scanning the column.
We would rather place them in the following part.
In the third step, we fill the rest of the bytes with the normal columns.
These columns can be divided into bytes and placed in arbitrary order.
In the following iterations, we generate the format of other parts using the same strategy.

\textbf{Design Trade-off by $th$.}
The underlying trade-off of setting $th$ is as follows: A larger $th$ ensures a highly effective PIM bandwidth as all the key columns are more compact.
However, this may split the table into too many parts.
CPU needs to access more cache lines to reform the row, leading to low effective CPU bandwidth.
This trade-off is validated through the experiments conducted in \sectionautorefname{}~\ref{sec:evaluation:format}.
The threshold is system- and workload-dependent, primarily influenced by transaction/analytical query rates and system bandwidth.
Specifically, if the workload is predominantly OLTP, a lower $th$ value can be selected to optimize CPU bandwidth.
Conversely, if the workload is predominantly OLAP, a higher $th$ value should be chosen to maximize PIM bandwidth.
This threshold often remains constant during runtime. Adjustment is typically unnecessary when query rates are stable. However, specific conditions may warrant threshold modifications. For instance, when CPU-side DRAM bandwidth significantly exceeds OLTP requirements, increasing the threshold can optimize OLAP performance. 

\textbf{Discussion on Key Column.}
Although for an actual HTAP database, all columns can be scanned with analytical queries, it does not mean that we need to conservatively regard all columns as key columns.
We can still perform analytical queries on normal columns that are distributed across devices through the CPU, albeit with a performance loss.
The indivisibility of key columns restricts the opportunity to generate a data layout with high effective bandwidth.
We justify this conclusion in \sectionautorefname{}~\ref{sec:evaluation:format}.
Therefore, in the actual deployment, we can prioritize the performance of frequent queries and choose fewer key columns.

Our architecture maintains compatibility with traditional methods for handling variable-width columns, though it does not specifically optimize for them. In practical implementations, variable-width columns are typically handled using traditional storage methods, such as length-prefixed encoding or separate metadata structures \cite{C-store10.1145/3226595.3226638}.

\subsection{Block-circulant Data Placement for PIM Parallelism}
\label{sec:layout:parallelism}

\begin{figure}
    \centering
    \includegraphics[width=\linewidth]{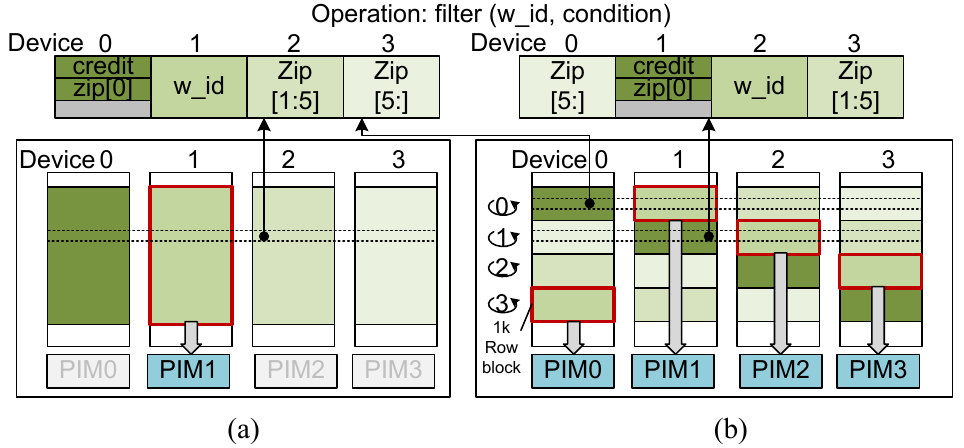}
    \vspace{-20pt}
    \caption{(a) Compact aligned format with low parallelism. (b) Block-circulant placement to fully exploit parallelism.}
    \label{fig:block_rotation}
    \Description{In this figure, we present the block-circulant data placement strategy. In subfigure (a), we present that the compact aligned format has low parallelism, because a whole column is mapped to one device. In subfigure (b), the block-circulant placement to fully exploit parallelism is presented. The table is divided into blocks across the rows along the IDE dimension, and each block is rotated to the next device. This block-circulant placement makes every column evenly distributed to all the PIMs, and fully utilizes the parallelism of PIM}
    \vspace{-10pt}
\end{figure}

With the compact alignment in the ADE dimension, one placement strategy for the IDE dimension is to align all the rows to the devices, as shown in \figurename{}~\ref{fig:block_rotation} (a).
However, the probability of being analyzed in OLAP workloads differs across the columns.
For example, eight queries analyze column \texttt{id}, while only three queries analyze column \texttt{state} \cite{CH-benchmark10.1145/1988842.1988850,tpch}.
Moreover, these two columns are analyzed by different queries, and it is hard to schedule the two analysis tasks in parallel.
Load imbalance emerges across PIM units as a ``hotspot'' column may be mapped to one PIM device.

We present a block-circulant data-placement strategy to fully exploit the parallelism of PIM, as shown in \figurename{}~\ref{fig:block_rotation} (b).
The table is divided into \emph{blocks} across the rows along the IDE dimension, and each block contains $B$ (suppose $B$=1024) rows of data.
In the first block (rows 0-1023), column $i (i=0, ..., 3)$ is mapped to device $i$.
Then, in the second block (rows 1024-2047), the columns are rotated, i.e., column $i$ is mapped to device $(i+1)\%4$.
Subsequent blocks perform the same rotation.
With block-circulant data placement, each column is evenly distributed to all the PIMs.
PIM parallelism can be fully utilized when scanning any column.

The block should be large enough to prevent the formation of too many discontinuous small blocks.
The block size should at least cover a row buffer of DRAM to ensure a relatively high row hit rate.
We set the block size to 1024.

\section{MVCC Support}
\label{sec:operations}

MVCC is essential for single-instance HTAP databases, and \archname{} supports MVCC.
For OLTP workload, \archname{} retains the same operational processes as traditional HTAP systems, with modifications only to the data storage format of MVCC. However, for OLAP workloads, we optimize the two MVCC operations, snapshotting and defragmentation, to minimize data communication overhead.

\subsection{MVCC Storage}
\label{sec:layout:mvcc}

\begin{figure}
    \centering
    \includegraphics[width=\linewidth]{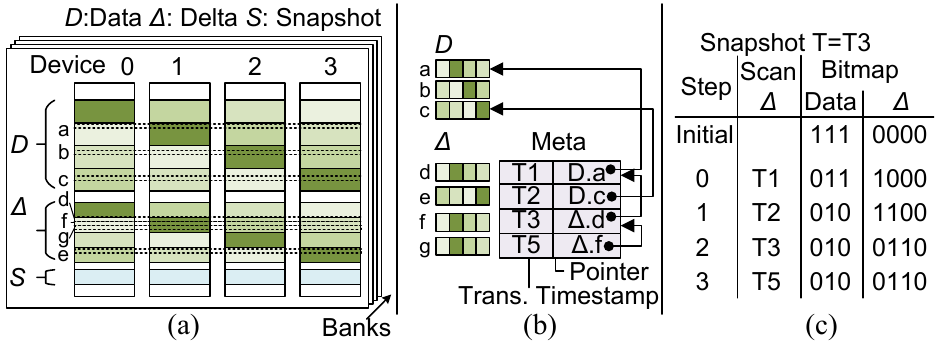}
    \vspace{-15pt}
    \caption{(a) Data region and delta region for MVCC. (b) The metadata format. (c) Snapshotting.}
    \label{fig:mvcc}
    \vspace{-10pt}
    \Description{This figure shows the MVCC storage and snapshotting. In subfigure (a), the data region and delta region for MVCC are presented. The data region contains the original version of rows, while the delta region contains the newer version of rows. In subfigure (b), the metadata format is presented. The metadata contains the transaction timestamps and the pointer to its last version row. In subfigure (c), the snapshotting procedure is presented. The snapshot is a collection of valid row pointers and is updated by scanning the version chain.}
\end{figure}

The data store format for supporting MVCC in \archname{} is shown in \figurename{}~\ref{fig:mvcc}~(a).
The storage of a table is divided into two regions, \emph{data region} and \emph{delta region}.
The data region contains rows of original versions, while newer versions created by transactions are stored in the delta region.
As metadata is not required by PIM units, it is stored in CPU memory.
The new versions of a row have the same rotation as its origin row in the data region, so that we can directly use PIM units to move the newest version back to the data region during defragmentation.
Therefore, the delta region is also organized into blocks.

\figureautorefname{}~\ref{fig:mvcc}~(b) depicts an example of version chain generated by transactions in \archname{}.
When a transaction T1 updates row $a$ in block 1, this row is right-rotated by 1 column, and therefore, its newer version should be put in block 1 of the delta region, which has the same rotation.
The CPU allocates an empty row $d$ in this block and records the transaction timestamps and the pointer to the origin row as metadata.
Therefore, the newer version's column is aligned to its origin row. 
If a transaction updates a row that already has a new version, for example, transaction T3, the CPU allocates an empty row $f$ in the same block.
The pointer points to the most recent version $d$ and forms a version chain.

\subsection{Snapshotting}
\label{sec:operations:snapshotting}

Snapshotting is executed before analytical queries to make PIM units operate on the rows of the correct version.
In a traditional database, the snapshot is a collection of valid row pointers and is updated before every analytical query.
We adopt a similar strategy to maintain the snapshots for \archname{}.
However, in \archname{}, each row is distributed across multiple devices.
Therefore, each device should maintain a copy of the snapshot.
To reduce the snapshot storage overhead, we encode the snapshots into bitmaps as the address of each row is relatively fixed in \archname{}.
The snapshot contains two bitmaps containing the visible information of the data and delta regions, respectively.
Each bit in the bitmap indicates the visible state of a row in the snapshot.
For example, bit `1' in the $i^\mathrm{th}$ position indicates the $i^\mathrm{th}$ row is visible in the snapshot, 
while bit `0' indicates that the row is invisible.
To minimize the data communication between CPU and PIM units, we maintain a dedicated region in memory banks to store the bitmap of rows in this bank, as shown in \figurename{}~\ref{fig:mvcc} (a).
In the proposed unified data format, each row is distributed across the devices in a bank; therefore, each device in this bank should store one copy of this bitmap.
During snapshotting, CPU updates the bitmap according to the metadata.
The bitmaps in these devices are also aligned across the ADE dimension so that CPU can update them simultaneously.

An example of snapshotting is shown in \figurename{}~\ref{fig:mvcc} (c).
Suppose at time \texttt{T4}, we start to execute one analytical query, and the last query is issued at \texttt{T0}.
Transactions \texttt{T1}, \texttt{T2}, and \texttt{T3} are generated after the last analytical query; therefore, these three transactions have not been updated to the bitmap.
During snapshotting, we need to update the bitmap according to the metadata one after another.
\texttt{T1} updates row $a$ with new version stored in row $d$.
Therefore, the bit related to row $a$ is set to `0', and the bit related to row $d$ is set to `1'.
Scanning \texttt{T2} and \texttt{T3} have the same operation as scanning \texttt{T1}.
T5 is generated after this analytical query is issued; therefore, it is skipped during snapshotting.

\subsection{Defragmentation.}
\label{sec:operations:defragmentation}

After a certain period, the transactions have updated numerous rows, and the original version of these rows in the data region is no longer used.
To clean up these outdated rows, \archname{} performs defragmentation periodically.
The newest version rows in the delta region are moved to the data region and overwrite their origin rows.
OLTP is paused during defragmentation to avoid data contention.

There are two candidate strategies to process the data movement.
The first strategy is to move data with the CPU, which has two steps:
(1) CPU reads the metadata from DRAM and merges it. (2) CPU processes the data movement according to the metadata. 
Due to the low bandwidth of the memory bus, copying rows is inefficient for tables with large row widths.
The second strategy is to move with PIM units.
Due to the data format in \archname{}, each row is distributed across the devices.
Therefore, the pointer field of the metadata should be broadcast to these devices so that PIM units in every device know where to copy the data.
This strategy has three steps:
(1) CPU reads out the metadata from DRAM. (2) CPU broadcasts the metadata to the devices. (3) PIM units merge the metadata and copy the new versioned data according to the metadata. 
Although this strategy can utilize the high bandwidth of PIM units, broadcasting the metadata involves additional communication.
Therefore, this strategy is suitable when the row width is much larger than the metadata size.

We can quantify the communication overhead to apply different defragmentation strategies according to the table's row width.
Suppose the DRAM rank has $d$ devices, the row width is $w$, and the metadata has $m$ bytes.
The delta region has $n$ rows, and $p$ of these rows are the newest version and need to be copied back to the data region.
CPU memory bandwidth is $bdw_{CPU}$, and the summation of PIM units bandwidth is $bdw_{PIM}$.
Therefore, the metadata has a total of $mn$ bytes.
$np$ rows of $dw$ bytes, a total of $npdw$ bytes, are moved from the delta region to the data region.
For the first strategy, the overall communication overhead is:
\begin{eqnarray}
    comm_{CPU} & = & \dfrac{mn + 2npdw}{bdw_{CPU}}
    \label{eqn:update-cpu}
\end{eqnarray}
The first term indicates reading the metadata, and the second term is the data movement overhead.
The overall communication overhead of the second strategy is:
\begin{eqnarray}
    comm_{PIM} & = & \dfrac{mn+dmn}{bdw_{CPU}} + \dfrac{dmn+2npdw}{bdw_{PIM}}
    \label{eqn:update-pim}
\end{eqnarray}
The first term indicates CPU reading out the metadata, and the second term is CPU broadcasting them.
The third term is PIM units reading metadata, and the fourth term is PIM units moving the rows.
From Equation.~\ref{eqn:update-cpu} and \ref{eqn:update-pim}, the second strategy is better when:
\begin{eqnarray}
    w & > & \dfrac{bdw_{PIM}+bdw_{CPU}}{2p\cdot(bdw_{PIM}-bdw_{CPU})} \cdot m
    \label{eqn:choose-update-with-pim-or-cpu}
\end{eqnarray}
For example, suppose $m=16$, $p\approx 1$, and $bdw_{PIM}:bdw_{CPU}=3:1$, the defragmentation is better to be executed with PIM units when $w>16$.

\section{Architecture Support}
\label{sec:architecture}

\subsection{Architecture Overview}
\label{sec:architecture:overview}

\begin{figure}
    \centering
    \includegraphics[width=0.95\linewidth]{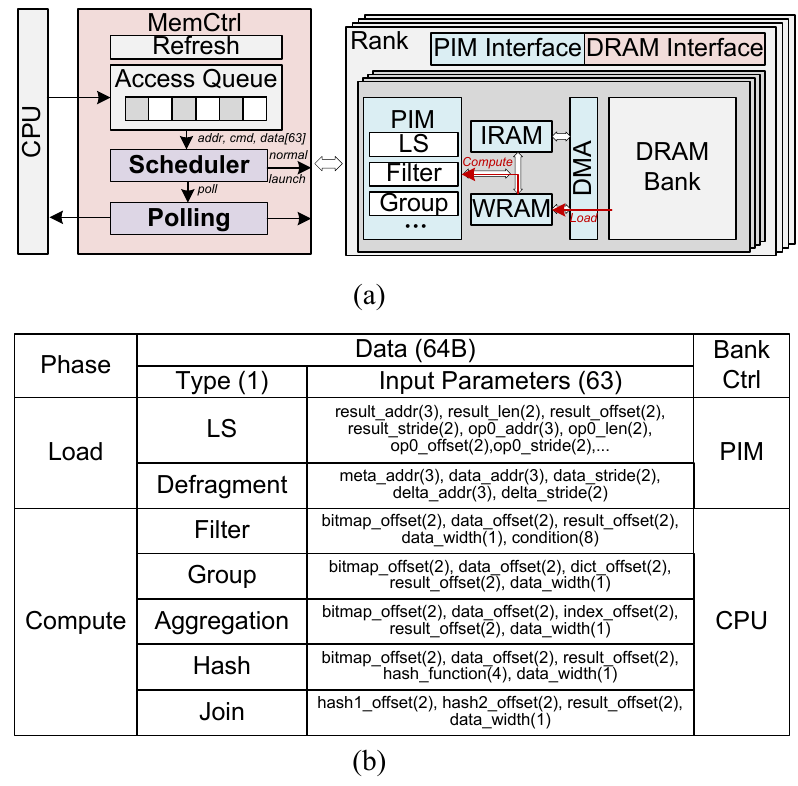}

    \vspace{-5pt}
    \caption{(a) Architecture. (b) Data fields of launch requests.}
    \label{fig:architecture}
    \vspace{-15pt}
    \Description{This figure shows the \archname{} architecture. In subfigure (a), we present the \archname{} architecture, which is an extension of the commercial general-purpose PIM architecture. Both of the two new components, the scheduler and the polling module, are added to the memory controller. We also present that in the load phase, PIM units read the data from DRAM banks. In the compute phase, PIM units perform the computation on the buffer, and the CPU can access the DRAM. In subfigure (b), we present the data fields of the launch requests. The launch request contains the operation type and input parameters. The field type occupies 1B while input parameters occupy 63 B. Load phase has two types of launch requests, \texttt{LS} and \texttt{Defragment}. \texttt{LS}'s input parameters is result\_addr(3B), result\_len(2B), result\_offset(2B), result\_stride(2B),op0\_addr(3B), op0\_len(2B), op0\_offset(2B),op0\_stride(2B). \texttt{Defragment}'s input parameters is meta\_addr(3B), data\_addr(3B), data\_stride(2B), delta\_addr(3B), delta\_stride(2B). And compute phase have 5 type of launch request, \texttt{Filter}, \texttt{Group}, \texttt{Hash}, \texttt{Join}, \texttt{Aggregration}. \texttt{Filter}, \texttt{Group}, \texttt{Hash}, \texttt{Aggregration} have common parameters, input bitmap\_offset(2B), data\_offset(2B), result\_offset(2B), data\_width(1B). The four requests also have their own parameters, e.g., Filter has 8B condition field. While Join have hash1\_offset(2B), hash2\_offset(2B), result\_offset(2B), data\_width(1B). This figure also shows that in the load phase, Bank control is on PIM while in the Compute phase, the bank control is on CPU}
\end{figure}

To support OLAP operations, and its concurrency with normal CPU access in OLTP workload, we extend the memory controller of commercial general-purpose PIM architecture \cite{UPMEM8875680} with two additional components, polling module and scheduler, as shown in \figurename{}~\ref{fig:architecture} (a).
The scheduler is responsible for recognizing the PIM unit control requests and orchestrating these requests and normal CPU access.
The polling module is responsible for automatically polling the PIM units and returning the PIM unit's finish signal to CPU.
With the help of the two modules, CPU only sends one message to the DRAM controller instead of to every PIM unit when controlling PIM units.
This reduces the communication overhead of each PIM task offloading and improves DRAM bandwidth utilization when CPU and PIM alternatively access DRAM.

There are two types of PIM unit control requests to be issued from CPU to execute OLAP operations, \emph{launch} and \emph{poll}.
These requests are disguised as normal memory accesses to a special physical address.
This special address is chosen from the unused DRAM address range and is preconfigured during the system boot.
The launch request is disguised as a memory write.
The data for writing contains the operation type and input parameters, summarized in \figurename{}~\ref{fig:architecture} (b).
The poll request is disguised as a memory read.
The scheduler can recognize the two requests according to their addresses and access type.
When recognizing a launch request, the scheduler broadcasts the operation type and input parameters to PIM units and launches PIM units by operating on the PIM interface \cite{UPMEM-Toolchain}.
Note that for general-purpose DRAM-based PIM \cite{UPMEM8875680}, the launch procedure has two main steps: handing over the control of the DRAM bank to PIM units and booting the PIM units.
In \archname{}, the scheduler only hands over the DRAM bank control to PIM units when the operation type is \texttt{LS} and \texttt{Defragment}, as other operations are processed on WRAM and do not require DRAM access.
To process a poll request, the scheduler notifies the polling module to poll the PIM units.
After all PIM units are finished, it returns a message to CPU through the DRAM read protocol.

\textbf{Discussion on Architecture Versatility.} 
The architecture demonstrates significant flexibility through the design of the launch request protocol. 
This adaptability stems from two key features: 
First, the CPU possesses complete configuration control over data fields, allowing tailored setups for different operations. 
Second, PIM units maintain programmability to interpret these customized fields. This dual-level configurability enables the architecture to support diverse scenarios, including AI and other mixed PIM-CPU tasks.

\subsection{OLAP Operations Execution}
\label{sec:architecture:execution_model}

The operations described in \figurename{}~\ref{fig:architecture} (b) are single-column operations processed by PIM units.
In this section, we present the execution of these operations.
In the conventional PIM program, the compute and load instructions are intertwined \cite{UPMEM-Toolchain}.
However, during the entire offloading process, the CPU's normal access is blocked, even when PIM units are executing computing instructions instead of loading data from DRAM, which does not efficiently utilize the available DRAM bandwidth.
To improve the bandwidth utilization, we present a \emph{two-phase execution} for OLAP operations.
Each OLAP operations are split into \emph{load phase} and \emph{compute phase}, and PIM units alternatively execute the two phases until the entire column is processed.
We take the filter operation as an example to present the two phases.
In the load phase, the CPU program sends a launch request with operation type \texttt{LS}.
DRAM bank access control is handed over to PIM units, and normal access from CPU is blocked.
According to the input parameter, PIM units store the results of the last compute phase that are buffered in WRAM (offset = \texttt{result\_offset}) back to DRAM and load new data to WRAM for the next compute phase.
Note that we use the block-circulant data placement described in \sectionautorefname{}~\ref{sec:layout:parallelism}, the real DRAM address of the data \texttt{op0} loaded by PIM unit $i$ is calculated by \texttt{op0\_stride}$*i +$ \texttt{op0\_addr}.
The WRAM size is 64 kB in \cite{UPMEM8875680} configuration, and the WRAM also serves as the operating memory for PIM units.
Therefore, we use only half of the WRAM (32 kB) to store the data.
According to our evaluation, it only takes ~300 $\mathrm{\mu}$s to load the 32kB data, meaning that the CPU normal memory access is blocked for no more than this period.
This blocking time is short enough for most second- and microsecond-level real-time OLTP databases \cite{Druid10.1145/2588555.2595631,sqlserver10.14778/2824032.2824071}.
In the compute phase, CPU sends a compute request of type \texttt{Filter}.
DRAM bank access control is not handed over to PIM units, and CPU can normally access the DRAM and execute transactions.
PIM units perform the filtering operation on the loaded data and store the result in the WRAM.
With the two-phase execution model, the normal access from CPU is not blocked when PIM units perform the computation and can better utilize the DRAM bandwidth.

\subsection{APIs for OLTP and OLAP Operations}
\label{sec:architecture:api}

\textbf{Transaction Commit.}
In \archname{}, the data in DRAM should be updated in time to ensure freshness for the OLAP workloads.
Therefore, we insert additional \texttt{clflush} instructions on the rows and memory barriers at the end of commits.

\textbf{Analytical Queries.}
\archname{} provides a set of APIs to support the analytical queries, including filter, aggregation, and hash join.
The last two operations are multi-column operations and require cooperation between the CPU and PIM units.
For aggregation with a form of \texttt{SUM(col1) GROUPBY col2}, the PIM units first execute the \texttt{Group} operation to scan column \texttt{col2} and compute indices of each data.
Then, CPU transfers the indices to the bank that stores the corresponding segment of column \texttt{col1} and launches PIM units to perform \texttt{Aggregation} operation.
For the hash join of two index columns, we adopt the same task division in \cite{upmem_join10.1145/3589258}.
PIM units first compute the hash value of the two columns with \texttt{Hash} operation.
After that, CPU fetches the hash values, divides them into buckets, and transfers them to the PIM units.
Finally, PIM units perform \texttt{Join} operation in their own buckets and get the join results.
In multi-column queries, columns are scanned serially, with PIM parallelism fully utilized during each scan due to block-circulant placement (\sectionautorefname{}~\ref{sec:layout:parallelism}).
Databases typically have sufficient rows to balance load across PIM units, ensuring efficient resource utilization.

\textbf{Data Re-layout.}
Byte-level re-layout is a common operation in current PIM systems, typically handled by the PIM runtime \cite{UPMEM-Toolchain}.
In \archname{}, we modify this function to support our data format.
This function takes three inputs: the data format information of tables, the row index, and the data buffer.
The data re-layout function is only invoked when 1) loading data from DRAM and 2) pushing the modified row back to memory during the transaction commit.
After data is loaded, it is stored in cache in its original format, allowing CPU to perform transactions on it directly.
This ensures that data re-layout only occurs when necessary, minimizing overhead and maintaining system efficiency.

\section{Evaluation}
\label{sec:evaluation}

\subsection{Experimental Setup}
\label{sec:evaluation:setup}

\textbf{Benchmarks.}
We evaluate \archname{} with CH-benchmark \cite{CH-benchmark10.1145/1988842.1988850}, which is a combination of two prominent database benchmarks, TPC-C \cite{tpcc} and TPC-H \cite{tpch}.
TPC-C is designed to measure OLTP performance by simulating a wholesale distribution business consisting of nine tables.
We simulate two transaction types, \emph{Payment} and \emph{New order}, which account for approximately 90\% of the TPC-C workload.
We execute transactions based on an open-sourced database, DBx1000 \cite{dbx100010.14778/2735508.2735511}, which supports the TPC-C workload and implements the MVCC scheme with row-store format.
To evaluate the performance of the column store and the unified data format in \archname{}, we extend DBx1000 with the corresponding data format.
TPC-H is an OLAP workload that evaluates data warehousing performance through complex queries.
We select three analytical queries from TPC-H for evaluation: aggregation-heavy query Q1, selection-heavy query Q6, and join-heavy query Q9.
The three analytical queries are chosen to represent different types of analytical workloads.
We implement the three queries and include both the PIM and CPU overheads described in \sectionautorefname{}~\ref{sec:architecture:api}.
The row number of table \texttt{ITEM}, \texttt{STOCK}, \texttt{CUSTOMER}, \texttt{ORDER}, \texttt{ORDERLINE}, \texttt{NEWORDER}, and \texttt{HISTORY} is set to 20M, 20M, 6M, 6M, 60M, 60M, and 6M, respectively.
The tables occupy 20 GB of memory storage.
The queries are scheduled using the method described in \cite{volcano10.14778/2002938.2002940}.
We use the hash index in DBX1000 to speed up the transaction and snapshotting during analytical queries.

\begin{table}
        \centering
        \caption{System Configuration}
        \label{tab:configure}

        \resizebox{\columnwidth}{!}{
        \begin{tabular}{cl}
                \toprule
                \multicolumn{2}{c}{Host CPU }                                                            \\\midrule
                Processor                       & 16 \texttimes{} O3CPU @3.2GHz                                          \\
                L1I/L1D                         & 32kB / 32kB,  Assoc: 8                                 \\
                L2 / L3                         & 1MB Assoc: 16 / 22MB, Assoc: 22                        \\
                Cache Line                      & 64 B                                                   \\
                \bottomrule
                \toprule
                \multicolumn{2}{c}{ DRAM DIMM }                                                          \\\midrule
                DRAM                            & DDR5-3200, 8$\times$8, 8GB/Rank                        \\
                Ba / De / Ro / Co               & 8 / 8 / 131072 / 1024                                  \\
                Timing Param.                   & tBURST=2.5ns  tRCD=tCL=tRP=7.5ns                    \\
                                                & tRAS=16.3ns tRRD=2.5ns                    \\
                                                & tRFC=121.9ns tWR=15.0ns tWTR=11.2ns                 \\
                                                & tRTP=3.75ns tRTW=tCS=4.4ns tREFI=3.9us                \\\bottomrule
                \toprule
                \multicolumn{2}{c}{ PIM Units }                                                          \\\midrule
                PIM Unit                        & 500MHz, 16 tasklets, 1GB/s bandwidth \cite{UPMEM8875680} \\
                                                & 64kB WRAM, 64-bit PIM-DRAM wire width \\
                Num                             & 64 per Rank, at Bank level inside Devices              \\
                \bottomrule
                \toprule
                \multicolumn{2}{c}{ System Configuration }                                               \\\midrule
                CPU System                      & 4 Channels $\times$4 Ranks normal DRAM                        \\
                & 4 Channels $\times$4 Ranks with PIM units \\
                \bottomrule
                \midrule
                \toprule
                \multicolumn{2}{c}{ HBM-based System Configuration } \\\midrule
                Host CPU and PIM Units & Same as DIMM-based system \\
                PIM DRAM & 32 Channels with PIM units \\
                & HBM3-2Gbps, 8Gb/Bank \\
                Pch / Bg / Ba / Ro / Co & 2 / 4 / 4 / 32768 / 64 \\
                Timing Param. & tBURST=2.0ns  tRCD=tCL=tRP=3.5ns                    \\
                & tRAS=8.5ns tRRD=2.0ns                    \\
                & tRFC=175.0ns tWR=4.0ns tWTR=1.5ns                 \\
                & tRTP=1.0ns tRTW=tCS=1.5ns tREFI=2.0us                \\\bottomrule
                
        \end{tabular}
        }

    \vspace{-10pt}
\end{table}

\textbf{Simulation.} The performance of \archname{} and baselines is evaluated with the ramulator-pim \cite{ramulator-pim}.
We use the ``OOO core'' CPU model of the simulator.
We extend the ramulator-pim simulator with Ramulator2 \cite{luo2023ramulator20modernmodular} for DDR5 DRAM modeling.
A PIM unit frontend is integrated into Ramulator2 to support both single-bank and parallel access.
We add two address mappings to support the two-dimensional access from CPU and PIM units.
We extend the Ramulator2's memory controller model with the two additional modules described in \sectionautorefname{}~\ref{sec:architecture:overview}.
System configuration is summarized in \tablename{}~\ref{tab:configure}.
The latency of handing over the bank access control is set to 0.2 $\mathrm{\mu}$s per rank, which is measured on a real general-purpose DRAM-based PIM server with an Intel Xeon CPU of 3.2GHz \cite{UPMEM8875680, UPMEM-Toolchain}.
Half of the WRAM (32 KB) is used to store the temporary data during the load phase.
The hardware modules in the memory controller are derived through Synopsys Design Compiler with TSMC 90nm technology library at the frequency of 2.4GHz.

To ensure a fair comparison with prior work \cite{Polynesia9835628}, we extended \archname{} to support HBM in addition to the default DIMM-based implementation. The detailed configuration of the HBM-based system is provided in \tablename{}~\ref{tab:configure}. 
Compared to DIMM-based system, only the PIM DRAMs are replaced with HBMs.
The PIM units and CPU-side configuration are kept the same.
The bank number of the HBM-based system is the same as the DIMM-based system.
Note that only the workload-wise performance comparison (\sectionautorefname{}~\ref{sec:evaluation:oltp_olap}) contains the comparison with HBM-based systems.
Other experiments are all on the default DIMM-based system.

\subsection{Results on Unified Data Format}
\label{sec:evaluation:format}

\begin{figure}
    \centering
    \includegraphics[width=\linewidth]{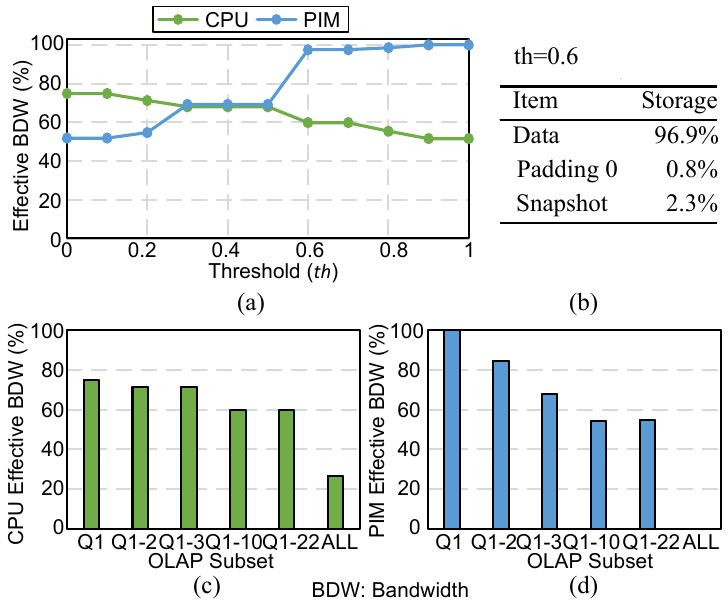}
    \vspace{-20pt}
    \caption{(a) CPU and PIM effective bandwidth under different $th$s. (b) Memory storage breakdown. (c) (d) Maximum CPU (PIM) effective bandwidth that ensures PIM (CPU) effective bandwidth $>70\%$, under different OLAP subsets. \emph{Q1-10} means the subset contains from \emph{Q1} to \emph{Q10}.}
    \label{fig:result_storage}
    \vspace{-5pt}
    \Description{This figure shows the performance of the compact data alignment technique. In subfigure (a), the x-axis is the $th$ value. The y-axis is the effective bandwidth of PIM and CPU. There are two lines, one is the PIM effective bandwidth and the other is the CPU effective bandwidth. With $th$ increasing from 0 to 1, the PIM effective bandwidth increases from around 55\% to 95\% while the CPU effective bandwidth decreases from around 75\% to 50\%. In subfigure (b), we present the memory storage breakdown. Data occupies 96.9\% of the memory storage. Padding `0's occupy 0.8\%. The snapshot bitmap occupies 2.3\% of the memory storage. In subfigures (c) and (d), both CPU and PIM effective bandwidth decrease when more queries are included in the workload subset.}
\end{figure}

We evaluate our unified data format with the CH-benchmark.
\figureautorefname~\ref{fig:result_storage}~(a) plots the effective bandwidth of both PIM and CPU under different values of the hyperparameter $th$, consistent with the trade-off analyzed in \sectionautorefname~\ref{sec:layout:alignment:compact}.
With $th=0$, the best CPU effective bandwidth of 74.8\% is achieved while PIM effective bandwidth is the lowest at 51.9\%.
When $th$ is set to $1$, PIM effective bandwidth is maximized while CPU effective bandwidth drops significantly.
To balance PIM and CPU bandwidth, we choose $th=0.6$ in our experiment.
At this value, PIM effective bandwidth reaches 97.4\%, while CPU effective bandwidth is 59.8\%, which is 15.0\% lower than the ideal bandwidth.
This effective bandwidth is sufficient for processing transactions, as they are compute-bound workloads.
Simultaneously, PIM can achieve a high effective bandwidth to ensure high OLAP performance.

The breakdown of the storage space is presented in \figureautorefname{}~\ref{fig:result_storage} (b).
The compact aligned format introduces negligible zero padding to the database storage, indicating its efficiency in space utilization.
The snapshot bitmap for supporting MVCC in \archname{} occupies only 2.3\% additional memory storage, demonstrating its minimal overhead.

To present the impact of key columns on the HTAP performance, we plot the maximum CPU (PIM) effective bandwidth at the minimum (maximum) $th$ value that ensures PIM (CPU) effective bandwidth $>70\%$, under different OLAP workload subsets, as shown in \figureautorefname{}~\ref{fig:result_storage} (c) and (d).
More queries in the workload subset lead to more columns being regarded as key columns.
For example, the subset \emph{Q1-1} contains only 4 key columns, while the subset \emph{Q1-3} contains 32 key columns.
\emph{ALL} represents all the columns are key columns, degraded to the na\"ive aligned format.
With more key columns, it becomes more difficult for both CPU and PIM units to achieve high effective bandwidth.
The maximum CPU effective bandwidth decreases from 74.8\% to 26.7\% when the workload subset increases from \emph{Q1-1} to \emph{ALL}.
The maximum PIM effective bandwidth decreases from 100\% to 54.7\%.
For \emph{ALL}, the CPU effective bandwidth never exceeds 70\%.
Therefore, in practice, it is better to select as few key columns as possible according to the OLAP workload.

To demonstrate the generality of our format algorithm, we also tested it on HTAPBench \cite{htap-mvcc10.1145/3514221.3526135}.
The results show that we achieve 57\%/98\% CPU/PIM bandwidth utilization when th=0.55 (not shown in the figure).

\subsection{OLTP and OLAP Performance}
\label{sec:evaluation:oltp_olap}

\begin{figure}
    \centering
    \includegraphics[width=0.9\linewidth]{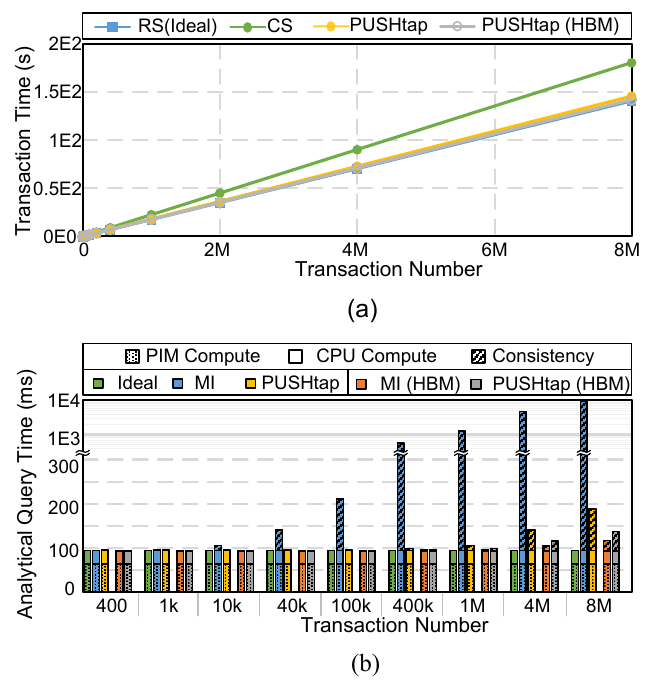}
    \vspace{-10pt}
    \caption{(a) Execution time of transactions with row-store (\emph{RS}), column-store (\emph{CS}) and \archname{}'s unified data format. (b) Analytical Query time breakdown of ideal, multi-instance (\emph{MI}) design and \archname{}, with different transaction numbers that update the data before the analytical query. The \emph{consistency} time includes the rebuilding (\emph{MI}) and snapshot \& defragmentation (\archname{}).}
    \vspace{-10pt}
    \label{fig:result_oltp_olap}
    \Description{In subfigure (a), the x-axis is the transaction number, from 0 to 8M. The y-axis is the execution time of transactions. There are four lines in the figure, representing row-store (ideal), column-store, \archname{} and \archname{} on HBM. All the lines increase linearly with the transaction number. Row store, \archname{} and \archname{} on HBM are close to each other. Column store has the longest execution time. In subfigure (b), the x-axis is the transaction number, from 400 to 8M. The y-axis is the execution time of analytical queries. The bars include Ideal, MI, \archname{} and \archname{} on HBM. MI's query time increases sharply when the transaction number is larger than 1M, much longer than all other bars. All three other bars are closed to ideal when the transaction number is smaller than 1M. Among the three bars, the query time of MI(HBM) is the shortest, while the query time of \archname{} is the longest.}
\end{figure}

\subsubsection{OLTP Performance}
\label{sec:evaluation:oltp_olap:oltp}

We first evaluate the OLTP performance of \archname{}, as shown in \figureautorefname{}~\ref{fig:result_oltp_olap} (a).
We compare \archname{}'s data format against row-store (\textbf{\emph{RS}}) and column-store (\textbf{\emph{CS}}) format on DIMM-based system.
The \emph{RS} Format is considered ideal for OLTP workloads as it aligns perfectly with their row-wise processing requirements.
In contrast, transactions using the \emph{CS} format require 28.1\% more execution time.
This is because the \emph{CS} format requires accessing data from every column to reconstruct the rows, leading to inefficiencies.
The analytical query latency is longer than \cite{Polynesia9835628} because our database scale is larger.
Therefore, more transactions can be executed during each analytical query.
In comparison, \archname{} only incurs a 3.5\% increase in execution time compared to \emph{RS}, which is attributed to the additional data re-layout operation.
This is because \archname{} can utilize the CPU's interleaving, as data is aligned along the ADE dimension.
Our compact aligned format optimizes this by splitting the data by bytes, allowing for compact arrangement.
The minor overhead observed in \archname{} is primarily due to the data reforming.

\textbf{OLTP Performance on HBM-based system.} Implementing \archname{} on HBM-based system (\emph{\archname{} (HBM)}) yields merely a 2.5\% speedup compared to DIMM-based system.
This marginal improvement stems from two fundamental mismatches:
First, the OLTP workload's non-memory-intensive nature fails to saturate HBM's high-bandwidth capabilities.
Second, the large interleave granularity of HBM necessitates the loading of more data per transaction (discussed in \sectionautorefname{}~\ref{sec:discussion}).
The bandwidth advantage of HBM is offset by the increased data transfer requirements.

\subsubsection{OLAP Performance}
\label{sec:evaluation:oltp_olap:olap}

We compare \archname{} with two DIMM-based baselines, \textbf{\emph{MI}} and \textbf{\emph{ideal}}, to present the OLAP performance.
\textbf{\emph{ideal}} assumes that all the columns are already compact, and the execution time only includes the scanning time.
\textbf{\emph{MI}} represents the multi-instance PIM-based HTAP system with data instances on both PIM and CPU memory space, utilizing suitable column-store and row-store formats, respectively~\cite{Polynesia9835628}.
To perform a fair comparison, we adapt the architecture from \cite{Polynesia9835628} to the same general-purpose DIMM-based PIM architecture as \archname{} (\figureautorefname{}~\ref{fig:architecture} (a)) while maintaining the same methodological approach.
Specifically, we replaced the PIM memory with the same DIMM DRAM modules used in \archname{}.
Instead of employing a dedicated hardware module for the rebuilding operations, we utilize the general-purpose PIM unit in the \emph{MI} system.
During the rebuilding stage, CPUs transfer all the new-versioned rows and corresponding metadata to DRAM banks, after which PIM units merge the metadata and copy the new-versioned data.
\archname{} generates a snapshot before every analytical query and ensures that PIM units skip old-versioned data when scanning the columns.
Defragmentation (presented in \sectionautorefname{}~\ref{sec:operations:defragmentation}) is executed after every 10k transactions.
This number is chosen based on the observation in \sectionautorefname{}~\ref{sec:evaluation:defragmentation}.
The result of analytical query execution time is plotted in \figureautorefname{}~\ref{fig:result_oltp_olap} (b).
Compared to \emph{ideal}, with 1M transactions, \emph{MI} introduces 123.3\% rebuilding overhead, while \archname{} incurs only 1.5\% overhead due to snapshot and defragmentation.
As the number of transactions increases (e.g., 1M), the rebuilding overhead in \emph{MI} causes a 13.3\texttimes{} slowdown in the analytical queries, significantly degrading system throughput.
In contrast, \archname{}'s snapshot and defragmentation overhead remains acceptable at 12.6\%.

\textbf{OLAP Performance on HBM-based system.} We also compare the performance of \archname{} with \emph{MI} on HBM-based system.
\textbf{\emph{MI (HBM)}} utilizes a dedicated rebuilding accelerator to perform the rebuilding operations, which is the same as \cite{Polynesia9835628}.
As the details of the dedicated rebuilding accelerator are not presented in \cite{Polynesia9835628}, we estimate the rebuilding overhead by the relative value to the CPU-based consistency, which is provided in \cite{Polynesia9835628}.
In contrast, \emph{\archname{} (HBM)}'s snapshotting and defragmentation are performed by CPU and general-purpose PIM units, which is the same as \archname{} on DIMM-based system.

The result is shown in \figureautorefname{}~\ref{fig:result_oltp_olap} (b).
Compared to \archname{} on DIMM-based system, \emph{\archname{} (HBM)} achieves 1.4\texttimes{} speedup when the transaction number is 8M, primarily due to a 2.1\texttimes{} reduction in defragmentation time thanks to HBM's high bandwidth.
With a dedicated rebuilding accelerator, \emph{MI (HBM)} (orange bar) only introduces 24.1\% rebuilding overhead, which is 4.1\texttimes{} lower than that of \archname{} on DIMM-based system, leading to a 1.6\texttimes{} increase in OLAP throughput.
However, the performance gains are relatively modest, allowing for the high cost of HBM and custom accelerators.
\archname{} on a general-purpose DIMM-based PIM architecture is a more cost-effective solution.

\begin{figure}
    \centering
    \includegraphics[width=0.75\linewidth]{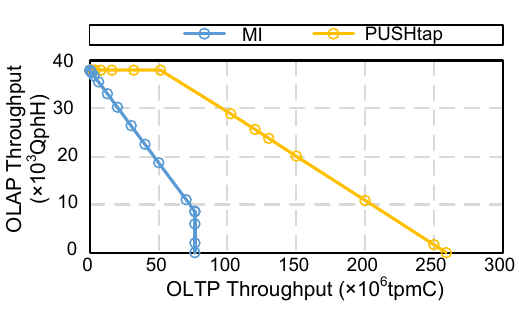}
    \vspace{-10pt}
    \caption{Throughput frontier for \emph{MI} and \archname{}.}
    \vspace{-10pt}
    \label{fig:result_htap}
    \Description{The frontier figure's x-axis is the OLTP throughput, and the y-axis is the OLAP throughput. The two lines are the frontier of \emph{MI} and \archname{}. The \archname{}'s frontier is shifted to the upper right.}
\end{figure}

\subsubsection{Performance Isolation}
\label{sec:evaluation:oltp_olap:throughput_frontier}

\figurename{}~\ref{fig:result_htap} plots the frontier \cite{howgoodhtapsystem10.1145/3514221.3526148} of OLTP and OLAP throughput for \emph{MI} and \archname{} on DIMM-based system.
Compared to \emph{MI}, the frontier of \archname{} is shifted to the upper right, indicating improved performance isolation due to the elimination of rebuilding overhead in single-instance design.
When the OLTP throughput of \archname{} $<$51.2M tpmC (transaction per minute, TPC-C), the OLAP throughput is constant at 38.0k QphH (query per hour, TPC-H), which is the peak OLAP throughput for \archname{}.
As the OLTP throughput increases, the OLAP throughput decreases because the memory system reaches the maximum overall bandwidth.
Compared to \emph{MI}, \archname{} can achieve 3.4\texttimes{} peak OLTP throughput.
When the OLTP throughput reaches 76.3 MtpmC, which is the peak value of \emph{MI}, \archname{} can still have 4.4\texttimes{} OLAP throughput, indicating better performance isolation.

\subsection{Defragmentation Operation}
\label{sec:evaluation:defragmentation}

\begin{figure}
    \centering
    \includegraphics[width=\linewidth]{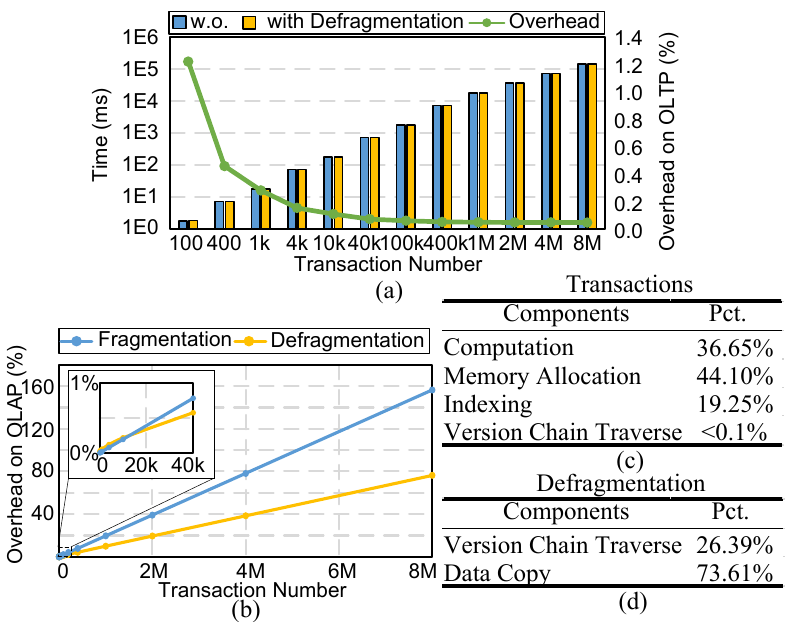}
    \vspace{-10pt}
    \caption{
        (a) Execution time of OLTP w/w.o. defragmentation.
        We additionally display defragmentation overhead on a secondary axis.
        (b) OLAP overhead caused by fragmentation. \emph{Fragmentation} is the performance degradation without defragmentation.
        \emph{Defragmentation} is the cost of periodic defragmentation.
        (c) (d) Breakdown of (c) transaction and (d) defragmentation. Fixed overhead is not included.
    }
    \vspace{-5pt}
    \label{fig:result_defregmentation}
    \Description{In subfigure (a), the x-axis is the transaction number, from 0 to 8M. The y-axis is the execution time of transactions. There are two bars in the figure, representing OLTP with and without defragmentation. The two bars are closed, with defragmentation overhead influence less than 1.5\% execution time. In subfigure (b), the x-axis is the transaction number, from 0 to 8M. The y-axis is the execution time of analytical queries. The fragmentation line is larger than the defragmentation line. In subfigures (c) and (d), we present the breakdown of transaction and defragmentation time. A transaction has 36.7\% computation, 44.1\% memory allocation, 19.3\% indexing, and less than 0.1\% traversing the version chain. A defragmentation has 26.4\% version chain traversing and 73.6\% data copy.}
\end{figure}

To show the necessity of defragmentation, we plot the defragmentation overhead and the performance degradation caused by fragmentation on OLAP in \figurename{}~\ref{fig:result_defregmentation}(b).
Without defragmentation, the analytical queries' execution time increases linearly with the transaction number, as more rows accumulate in the delta region.
The rows of old versions are skipped during analytical queries.
However, many row widths are smaller than 8 bytes, which is the minimum access granularity of PIM units \cite{UPMEM8875680}.
Skipping such discrete bytes does not save PIM bandwidth usage, and PIM units still load this unnecessary data.
As a result, fragmentation significantly degrades OLAP performance.
Periodically executing defragmentation is necessary to ensure OLAP performance.
When the transaction number exceeds 10k, the overhead caused by fragmentation is larger than the defragmentation overhead (2.05\texttimes{}).
This is because the fixed overhead, including thread creation and PIM units activation, is amortized when the number of transactions is large.
Therefore, we execute defragmentation every 10k transactions to minimize the overhead while preserving OLAP performance.

The defragmentation overhead on OLTP across varying transaction numbers is shown in \figurename{}~\ref{fig:result_defregmentation} (a).
It represents the ratio of defragmentation time to total transaction time.
In contrast with OLAP, the defragmentation only introduces $<$1.5\% overhead to OLTP.
\figurename{}~\ref{fig:result_defregmentation} (c) and (d) present the breakdown of transaction and defragmentation time.
Less than 0.1\% time of each transaction is spent on traversing the version chain, as only one version chain is traversed per transaction.
Most time is occupied by indexing, memory allocation, and computation.
Defragmentation on a row involves traversing the version chain and copying the data, which is negligible compared to a transaction.

\begin{figure}
    \centering
    \includegraphics[width=\linewidth]{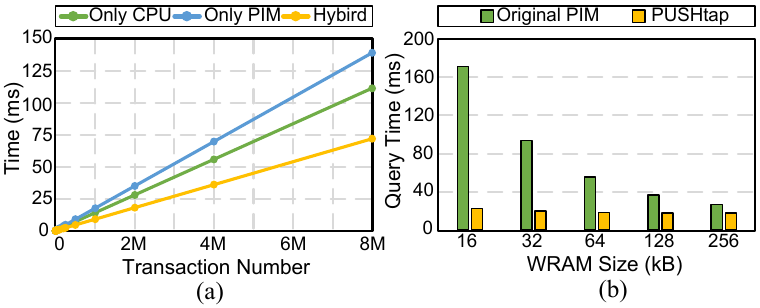}
    \vspace{-15pt}
    \caption{(a) Defragmentation time with (1) purely CPU, (2) purely PIM units, and (3) strategy in \sectionautorefname{}~\ref{sec:operations:defragmentation} (\emph{Hybrid}). (b) \emph{Q6} execution time across different WRAM sizes.}
    \vspace{-5pt}
    \label{fig:result_defreg_method_arch}
    \Description{In subfigure (a), the x-axis is the transaction number, from 0 to 8M. The y-axis is the defragmentation time. The three lines are the defragmentation time with CPU, PIM units and hybrid strategy. The CPU strategy has the longest defragmentation time, while the hybrid strategy has the shortest defragmentation time. In subfigure (b), the x-axis is the WRAM size, from 16kB to 256kB. The y-axis is the execution time of analytical queries. The bars include original PIM architecture and \archname{} on a DIMM-based system. The query time of the original PIM architecture decreases when the WRAM size increases. The query time of \archname{} is nearly constant when the WRAM size increases, with all the bars shorter than the original PIM architecture.}
\end{figure}

\figureautorefname{}~\ref{fig:result_defreg_method_arch} (a) plots the defragmentation time with (1) only CPU, (2) only PIM units, and (3) strategy presented in \sectionautorefname{}~\ref{sec:operations:defragmentation} (denoted as \emph{Hybrid}).
Neither CPU- nor PIM-only strategy can achieve optimal defragmentation efficiency.
With our unified data format, the table parts' row width varies from 2 bytes to over 20 bytes.
According to our conclusion in \sectionautorefname{}~\ref{sec:operations:defragmentation}, these parts are suitable for different strategies.
The \emph{hybrid} selects different strategies depending on the tables' row widths and can achieve the best efficiency.

\subsection{Architecture Comparison}


We compare the PIM architecture of \archname{} with the original general-purpose DRAM-based PIM architecture \cite{UPMEM8875680}.
Both architectures adopt the two-phase execution presented in \sectionautorefname{}~\ref{sec:architecture:execution_model}, with the only difference on the communication overhead presented in \sectionautorefname{}~\ref{sec:architecture:overview}.
\figureautorefname{}~\ref{fig:result_defreg_method_arch} (b) depicts the execution time under different WRAM sizes.
With a larger WRAM, fewer load phases are required, reducing the CPU-PIM mode switch overhead.
The execution time of the original PIM architecture decreases by 6.4\texttimes{} when increasing WRAM size from 16 kB to 256 kB, as the mode switching overhead drops from 88.8\% to 35.3\% of the computing time.
Accordingly, the period of the load phase increases to $>1$ ms when the WRAM size increases to 256 kB, limiting the usage of real-time OLTP.
The CPU-PIM mode switch overhead has minimal effect on the performance of \archname{}, as the mode switch function is offloaded to memory channels and only accounts for 7.0\% of the computing time on average.
For the default 64kB configuration, \archname{} can achieve 3.0\texttimes{} speedup compared to the original PIM architecture.

\subsection{Area Overhead.}
\label{sec:evaluation:area}
The additional hardware modules introduce minimal area overhead of 0.115 $mm^2$ in an 8-channel memory controller, with the scheduler occupying 0.112 $mm^2$ and the polling module requiring only 0.003 $mm^2$. This overhead is negligible compared to the total memory controller area of approximately 13 $mm^2$ \cite{SapphireRapids9731107}. 

\section{Discussion}
\label{sec:discussion}



\textbf{PIM Technique Selection.}
As memory interleaving is a well-established technique used in various memory systems, making it feasible to integrate PIM units into these existing systems \cite{AiMHW9731711, UPMEM8875680, HBM-PIM-HW9365862,smartSSD9141369}.
For PIM-based HTAP architecture with unified data format, the interleave granularity should be fine enough so that both CPU and PIM can efficiently access the small data elements in databases.
For instance, \texttt{ORDERLINE} table's \texttt{amount} column is only 8 bytes in size.
If the interleave granularity is set to 64 bytes, we should access an additional 48 bytes to acquire this 8-byte valid data, which results in bandwidth waste.

We compare three techniques --- HBM, DIMM, and SSD --- on their interleave granularity.
We take the CH-benchmark as an example \cite{CH-benchmark10.1145/1988842.1988850,tpch}.
Its column width varies from 2 bytes to 152 bytes.
SSD's interleave granularity is $\approx 1$ MB.
HBM provides a 64-byte (or 32-byte) granularity.
DIMM offers the finest granularity at 8 bytes.
DIMM's finest granularity allows us to optimize access waste through mapping methods.
Therefore, we choose DIMM DRAM as the PIM storage.

\textbf{Compact Aligned Format on Command-Driven PIM.}
The data format method employed in \archname{} is also well-suited for command-driven PIM architectures \cite{AiMHW9731711,kim2023darwin}.
These architectures typically require a high degree of customization of the database instruction set.
In such an architecture, we can design specialized accelerators like in \cite{Polynesia9835628} to expedite time-consuming operations, such as defragmentation and data layout, which were identified as performance bottlenecks in our evaluations.
By leveraging these accelerators, we can further reduce the performance gap between OLTP/OLAP operations and their ideal performance benchmarks.
This enhancement ultimately boosts overall system throughput.
We plan to explore this potential in future research to fully realize the benefits of these optimizations.

\vspace{-0.3cm}
\section{Related Work}

\textbf{PIM}. 
In PIM, PIM units can be located in various memory hierarchies, including cache \cite{affinity10411388,infinity-stream10.1145/3582016.3582032,near-stream9773240,dalorex10071089}, DRAM memory (DDR\cite{tensordimm10.1145/3352460.3358284,RecNMP9138955,UPMEM8875680,upmem_join10.1145/3589258,axdimm10.1145/3533737.3535093},GDDR\cite{AiMHW9731711}, HBM\cite{Polynesia9835628,HBM-PIM-HW9365862}), and SSD \cite{HTAP-DNP10.14778/3547305.3547307,smartSSD9141369}.
There is no direct connection between the PIM units distributed in sub-modules of memory; the communication has to go through the CPU, resulting in high-cost inter-PIM-unit communication.
Current works treat the PIM units as distributed systems.
They are devoted to static task division to ensure PIM load balance and low communication overhead \cite{graph_ndp7284059,baek1psyncpim,graphq10.1145/3352460.3358256,gearbox10.1145/3470496.3527402,cross-level-recommand10.1145/3579371.3589101, upmem_join10.1145/3589258}.
While others propose to add additional connections and access modes to realize an automatic communication and load balance \cite{abndp10.1145/3582016.3582026,dimmlink10071005,ndp-bridge,abc-dimm9499805}.

The distributed PIM units also face the conflict that memory interleaving prevents PIM units from being visible to a contiguous block of data.
Some workloads, e.g. element-wise operations, can coexist with interleaving as PIM units still have access to complete data elements \cite{tensordimm10.1145/3352460.3358284,chopim9138972,pidram10.1145/3563697}.
For a general-purpose scenario, some works\cite{UPMEM8875680,HBM-PIM-HW9365862,topim10.1145/3470496.3527431} divide a dedicated PIM memory space from the main memory and manually re-layout the data when writing to the space.
UM-PIM \cite{um-pim} proposes to use dynamic address mapping to enable two different memory pages with different data layouts to co-exist in the PIM system.
In this work, instead of resolving the conflict, we utilize PIM and memory interleaving to provide two-dimensional memory access.
We exploit its benefit in an appropriate scenario -- HTAP.

\textbf{HTAP Architecture. }
HTAP research mainly focuses on data storage format \cite{OracleDF7113373,sqlserver10.14778/2824032.2824071,TiDB,SAPHANA10.1145/2213836.2213946,singlestore}, query optimization \cite{thecaseitem_7e8d7bab152f4630a9d2d4e4412ff582,scheduling1_10.14778/3476311.3476378,howgoodoptimizer10.14778/2850583.2850594,rethink10.1145/3626750}, indexing technique \cite{mvpbtriegger2020mv,htap-mvcc10.1145/3514221.3526135}, scheduling \cite{SAPHANA10.1145/2213836.2213946,elasticscheduling10.1145/3318464.3389783,performancecharacterization9458644}, and accerleration with computational storage \cite{HTAP-DNP10.14778/3547305.3547307,morethanjust10.14778/3583140.3583161}.

\textbf{In-Memory Database.}
In-memory database is proposed to provide real-time transaction processing \cite{in-memory-db-review10.1145/2814710.2814717}.
Compute Express Link (CXL) technique is introduced to improve in-memory database's scalability \cite{cxl-db-eval10.1145/3533737.3535090} and to provide a new solution for durability \cite{cxl-db10597964}.
The CXL latency can be hidden by properly prefetching data to local memory or cache in OLTP workloads \cite{cxl-db-eval10.1145/3533737.3535090}.

\section{Conclusion}
\label{sec:conclusion}
This work proposes \archname{}, a PIM-based HTAP system with a unified data storage format tailored for both OLTP and OLAP workloads. 
By combining the access of PIM units and CPU, we create a two-dimensional access memory space. 
We introduce a unified data format to enhance effective bandwidth, ensure PIM load balance, and support MVCC. 
\archname{} also allows concurrent CPU and PIM access, optimizing HTAP performance.  
Extensive experiments show that \archname{} can deliver satisfactory performance gain.

\begin{acks}
We sincerely thank our shepherd, Prof. Onur Mutlu, and the anonymous reviewers for their comments and suggestions to improve the
paper. 
We also thank Huan Zhou for improving the figures.
\end{acks}

\bibliographystyle{ACM-Reference-Format}
\bibliography{ref}



\end{document}